\documentstyle[osa,aps,prb,epsfig]{revtex}
\begin{document}

\twocolumn[\hsize\textwidth\columnwidth\hsize\csname
      @twocolumnfalse\endcsname

\title{Abrupt metal-insulator transition observed in VO$_{2}$ thin films induced by
a switching voltage pulse}

\author{Byung-Gyu Chae, Hyun-Tak Kim, Doo-Hyeb Youn, and Kwang-Yong Kang}

\address{Basic Research Laboratory, ETRI, Daejeon 305-350,
Republic of Korea}

\maketitle{}

\begin{abstract}
  An abrupt metal-insulator transition (MIT) was observed in VO$_{2}$ thin films
during the application of a switching voltage pulse to
two-terminal devices. Any switching pulse over a threshold voltage
for the MIT of 7.1 V enabled the device material to transform
efficiently from an insulator to a metal. The characteristics of
the transformation were analyzed by considering both the delay
time and rise time of the measured current response. The
extrapolated switching time of the MIT decreased down to 9 ns as
the external load resistance decreased to zero. Observation of the
intrinsic switching time of the MIT in the correlated oxide films
is impossible because of the inhomogeneity of the material; both
the metallic state and an insulating state co-exist in the
measurement volume. This indicates that the intrinsic switching
time is in the order of less than a nanosecond. The high switching
speed might arise from a strong correlation effect (Coulomb
repulsion) between the electrons in the material.
\\
\\
\end{abstract}
\pacs{ }
]

  Vanadium dioxide, VO$_{2}$, possesses a first-order metal-insulator transition (MIT)
making it an attractive material for switching
devices.\cite{1,2,3} The MIT occurs near 68 $\rm^{\circ}$C and is
accompanied by a structural phase transition. Various transition
properties have been studied such as crystal structure and other
physical quantities. In particular, aspects of the transition have
been examined during thermal and optical
inducements.\cite{3,4,5,6,7} It is also known that a negative
differential resistance (NDR) is observed when the current-voltage
characteristic of this material is controlled by a static
current.\cite{8,9} Such an experiment allows the measurement of
the MIT with respect to temperature. The current-controlled NDR
has been widely investigated for the various compounds of vanadium
oxide.\cite{8,9,10} The resistance of the systems abruptly changes
at the transition point in contrast to the NDR properties of a
conventional semiconductor system. The MIT behavior controlled by
a static current is extremely stable and reversible.\cite{9,10} Up
until now, controversy over the mechanism of this transition has
existed. It is not known whether the transition is due to thermal
or electronic effects.

  Recent research favors the electronic model.
Observations of the sample stability and the current injected to
initiate the MIT support this view.\cite{9,10} We have reported a
stable MIT induced by a constant applied electric field in highly
oriented VO$_{2}$ films and revealed the mechanism of the MIT to
be based upon electron-electron correlation using a Raman study of
the planar devices.\cite{11} The transition speed of the MIT in
VO$_{2}$ films has been reported to be below a picosecond through
the use of ultrafast optical techniques.\cite{12} If the
field-induced MIT occurs quickly enough to apply to high speed
devices and shows a reproducible behavior, there are various
applications for VO$_{2}$ in switching devices such as electrical
switches, modulators, and electro-optical devices. Therefore, it
is very important to observe the time dependence of the
field-induced MIT and also to study the transient properties of
the MIT.

  In this letter, we investigate the transient current in the material during an abrupt MIT
induced by applying switching voltage pulses to VO$_{2}$-based
devices. VO$_{2}$ thin films with a preferential orientation of
(100) are used in these switching experiments. The transient
properties of the MIT are analyzed through the observation of the
current response profiles.

  (100) oriented VO$_{2}$ thin films were grown on $\alpha$-Al$_{2}$O$_{3}$ (1012)
by laser ablation. The partial pressure of oxygen during the
deposition process plays an important role in obtaining the pure
VO$_{2}$ phase. The VO$_{2}$ films were deposited in a working
pressure of 60 mTorr with the argon gas atmosphere containing 10\%
oxygen. The substrate temperature and the deposition rate of the
films were 450 $\rm^{\circ}$C and approximately 0.39 \AA/sec,
respectively. A detailed description of the conditions for the
deposition of the films were given in previous papers.\cite{13,14}

  VO$_{2}$-based two-terminal devices were fabricated using
well-established semiconductor techniques, as shown in the device
diagram of Fig. 1. The VO$_{2}$ films were isolated using
selective etching and Au/Cr electrodes were patterned on the films
using the lift-off method. The channel length and width between
two electrodes were 3 $\mu$m and 30 $\mu$m, respectively.

 A schematic of the apparatus used to observe the transient current profiles
is provided in Fig. 1. The system is similar to the Sawyer-Tower
circuit used for measuring the displacement current in
dielectrics. Switching voltage pulses are generated by a function
generator (HP 33120A) and pulse profiles are detected by a digital
oscilloscope (HP 54810A). The transient current profiles passing
through the device were observed via an external load resistance.
The current is calculated using the voltage traces measured over
the terminals of the load resistance and Ohm's Law.

  Figure 2 shows an abrupt MIT induced by a static electric field
applied to the VO$_{2}$ films. The film had an abrupt resistivity
change of an order of 3$\times$10$^{3}$ at a critical temperature
of $T_{c}$$\approx$338 K.\cite{11,13} A resistor of 1 k$\Omega$
was connected to the device in series to prevent an excessive
current-flow through the film, and the source-drain current
between two electrodes was measured, as shown in the inset to Fig.
2. When a static electric field is applied to the device, the
current slowly increases with increasing applied voltage and
finally jumps up to 5.4 mA at the transition voltage, $V_{MIT}$,
of 7.1 V. This behavior was described in detail in previous
papers.\cite{11,13}

  Figure 3 shows the transient current profiles measured while applying
an external switching voltage pulse to the two-terminal device.
The switching experiment was conducted at room temperature and
used a load resistance of 1 k$\Omega$. A single pulse with a width
of 1 $\mu$s was applied to the device. We observed that the peak
value in the measured current profiles abruptly increase when the
applied pulse exceeds the MIT threshold voltage of 7.1 V. For an
applied pulse of 7 V, which is below $V_{MIT}$, the height of
current profile remains at approximately 300 $\mu$A, as shown in
the inset to Fig. 3. This behavior is similar to that obtained
using a static applied voltage as shown in the current vs. voltage
measurement displayed in Fig. 2. At a peak voltage of 10 V, the
current reaches a value of 7.5 mA. This corresponds to a current
density of 2.5$\times$10$^{5}$A/cm$^{2}$, which is of the order of
that in a dirty metal. This is a MIT induced by a switching
voltage pulse.

  In order to explore the switching speed of the MIT for a VO$_{2}$ film,
the current profiles were measured by varying the external load resistances in the circuit. Figure
4 displays the current profiles as a function of the load
resistance for measurements employing an applied voltage pulse of
10 V. The lower load resistance leads to the larger current flow
through a film after the transition.

  Figure 5(a) displays the current profiles observed using a 3 k$\Omega$
load resistance. The inset to the figure is a magnified view of
the curves in the vicinity of the transition region. There is a
time delay just before the current increases. We find that the
time delay arises during the increasing portion of the applied
switching voltage pulse. That is, although the applied pulse
increases steeply, there is a finite time before it reaches the
maximum voltage. As shown in the inset of Fig. 5(b), when the
applied pulse exceeds $V_{MIT}$ and approaches the maximum value,
the current profile begins to increase as indicated by an arrow. A
pulse voltage above $V_{MIT}$ induces the MIT. In contrast, a
delay is not observed at the falling edge of the current profiles,
as shown in the inset to Fig. 3. This is explained by the fact
that the applied pulse drops promptly through the value of
$V_{MIT}$ in the falling edge. We define the delay time,
$\Delta$$\tau$, as the interval between the onset of the applied
switching pulse and the corresponding onset of the current signal.
The linear extrapolation technique is used to determine the
starting point of the transition. Both the voltage and the current
signals are extrapolated to the lines where the current and the
voltage are zero. The delay time is estimated to be approximately
20 ns and has no clear dependence on the load resistance, as shown
in Fig. 5(b). Here, we did not consider the current profiles
observed using less than a 1 k$\Omega$ load resistance since the
resulting large current of approximately 10 mA can deform the
profiles. The rise time of an applied pulse can be reduced by
altering the experimental situation because it is related to the
capacitance of the pulse generator. If we can make an applied
pulse without a rise time, the delay time will approach zero.

  We find that the switching speed of the MIT corresponds to the rise time of the current profile.
Here, the rise time, $\tau_{R}$, is defined as the necessary for
the current increase to 90\% of its maximum value. A plot of
$\tau_{R}$ versus load resistance is presented in Fig. 5(b). The
plot is fitted with a straight line. When a $RC$ component is
included in the sample, $\tau_{R}$ can be expressed as $\tau_{R} =
\tau_{0} + RC$, where $\tau_{0}$ is the intrinsic rise time when
the resistance component is zero and $C$ is the capacitance of the
film. The rise time on the straight line when $R = 0$ is the
intrinsic rise time of the MIT, which is estimated to be 9 ns. It
is known that the MIT in a VO$_{2}$ film occurs with inhomogeneity
during the transition process.\cite{15,16} An intrinsic resistance
component inevitably exists in the films due to the spatially
inhomogeneous system which contains both the metallic state and an
insulating state co-existing in the measurement volume. Therefore,
it is impossible to observe the intrinsic switching time of the
MIT because of the remnant component of resistance. This indicates
that the intrinsic switching speed of the MIT in VO$_{2}$ films
can be lower than order of a nanosecond.

  The switching time for the MIT can be estimated from a simple thermal model
using the heat balance equation.\cite{9} Here the heat power is
assumed to be conserved in the device volume during the transition
process. The estimated value of the switching time is of the order
of a microsecond for a device of our scale. The thermal model does
not account for a transition time for the MIT of order of a
nanosecond. As one of the possible explanations, it may be
suggested that conducting paths are formed in the device volume.
The formation of conducting paths over repeated switching cycles
could lead to the MIT that is unstable and irreversible. However,
our experiments demonstrated that the transition was repeatable
and reversible. We suggest that the high speed switching is
attributable to a strong electron correlation effect as proposed
in Mott's construction of the metal-insulator transition.\cite{17}

  In conclusion, the abrupt MIT for VO$_{2}$-based two-terminal devices was observed
during the application of a switching voltage pulse. Current
profiles resulted from the MIT were measured using switching
pulses with a peak value above the MIT threshold voltage. The
delay time and the rise time were compared and the intrinsic
switching time was estimated to be about 9 ns. These planar
devices are thus identified as being highly applicable to various
switching systems because of their excellent switching behavior
and stability.

\begin{figure}
\vspace{-3.0cm}
\centerline{\epsfysize=14cm\epsfxsize=8.0cm\epsfbox{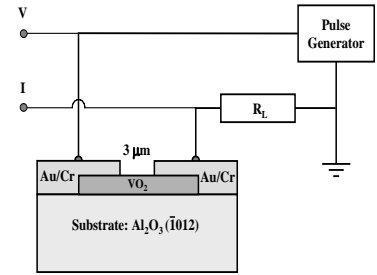}}
\vspace{-7.0cm} \caption{Schematic of the test circuit used for
monitoring the MIT in VO$_{2}$ thin films induced by a voltage
pulse.} \label{f1}
\end{figure}

\begin{figure}
\vspace{2cm}
\centerline{\epsfysize=14cm\epsfxsize=10cm\epsfbox{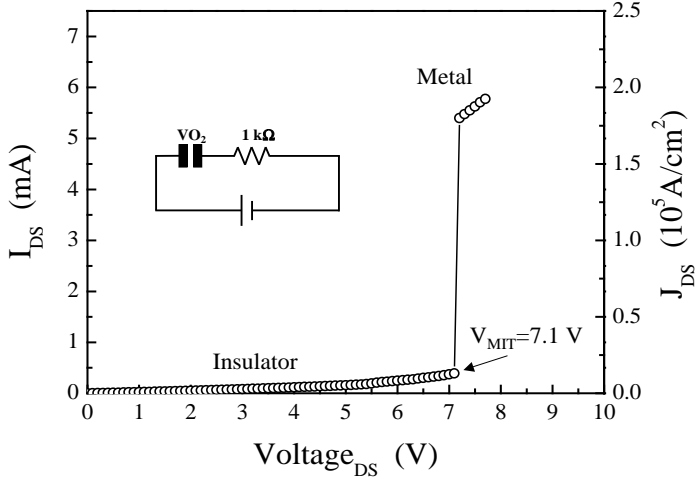}}
\vspace{-5.0cm} \caption{Abrupt metal-insulator transition induced
by a static electric field in a VO$_{2}$ thin film. The inset
displays the circuit used for experiments. I$_{DS}$ is the
drain-source current and J$_{DS}$ is the corresponding current
density.} \label{f2}
\end{figure}

\begin{figure}
\vspace{2.0cm}
\centerline{\epsfysize=12cm\epsfxsize=8cm\epsfbox{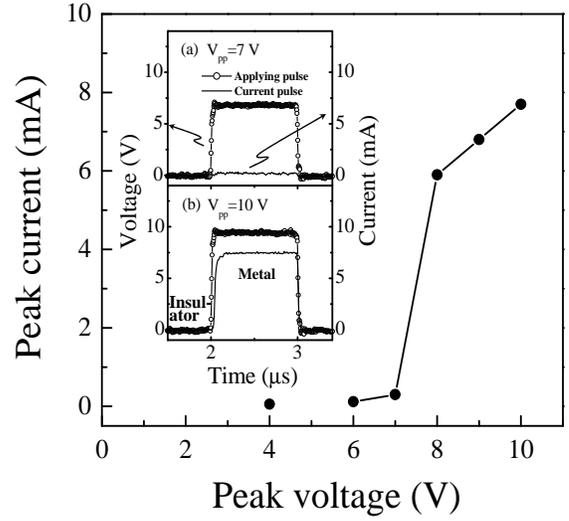}}
\vspace{-3.0cm} \caption{Peak current as a function of the peak
voltage of the switching pulse. The inset to the figure
illustrates the pulse profiles measured using applied switching
pulses of (a) 7 V and (b) 10 V.} \label{f3}
\end{figure}

\begin{figure}
\vspace{-0.3cm}
\centerline{\epsfysize=14cm\epsfxsize=10cm\epsfbox{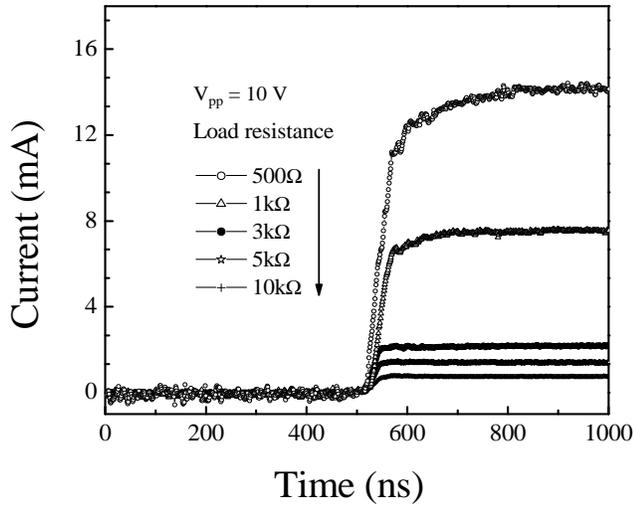}}
\vspace{-1.3cm} \caption{Current profiles from VO$_{2}$ devices
with varying load resistance.} \label{f4}
\end{figure}

\begin{figure}
\vspace{-0.5cm}
\centerline{\epsfysize=15cm\epsfxsize=10cm\epsfbox{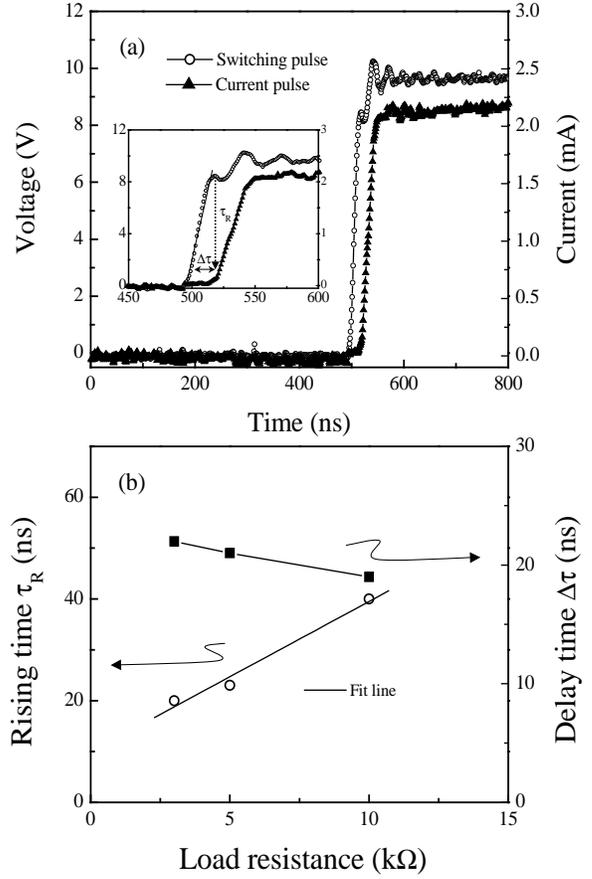}}
\vspace{-1.5cm} \caption{(a) Current profiles measured with a 3
k$\Omega$ load resistance. The inset to the figure gives a
magnified view of profiles. (b) Rise time and delay time as a
function of load resistance.} \label{f5}
\end{figure}

\end{document}